\documentclass[aps,prd,10pt,twocolumn,superscriptaddress,nobibnotes,longbibliography]{revtex4-1}
\pdfoutput=1

\usepackage[nodayofweek]{datetime} 
\usepackage{xcolor}
\newdateformat{daymonthyeardate}{\THEDAY\;\monthname[\THEMONTH], \THEYEAR}
\usepackage{mathtools}
\usepackage{natbib}
\usepackage{dcolumn}
\usepackage{bm}

\usepackage[utf8x]{inputenc}
\usepackage[T1]{fontenc}
\usepackage{enumitem}

\usepackage{amsmath}
\usepackage{amssymb}
\usepackage{amsfonts}
\usepackage{graphicx}

\usepackage[colorlinks=true, allcolors=blue]{hyperref}

\begin{document}

\title{Not as Big as a Barn: Upper Bounds on Dark Matter-Nucleus Cross Sections}
\author{Matthew C. Digman}
\email{digman.12@osu.edu}

\affiliation{Center for Cosmology and AstroParticle Physics (CCAPP), Ohio State University, Columbus, Ohio 43210, USA}
\affiliation{Department of Physics, Ohio State University, Columbus, Ohio 43210, USA}
\author{Christopher V. Cappiello}

\affiliation{Center for Cosmology and AstroParticle Physics (CCAPP), Ohio State University, Columbus, Ohio 43210, USA}
\affiliation{Department of Physics, Ohio State University, Columbus, Ohio 43210, USA}
\author{John~F.~Beacom}

\affiliation{Center for Cosmology and AstroParticle Physics (CCAPP), Ohio State University, Columbus, Ohio 43210, USA}
\affiliation{Department of Physics, Ohio State University, Columbus, Ohio 43210, USA}
\affiliation{Department of Astronomy, Ohio State University, Columbus, Ohio 43210, USA}
\author{Christopher M. Hirata}

\affiliation{Center for Cosmology and AstroParticle Physics (CCAPP), Ohio State University, Columbus, Ohio 43210, USA}
\affiliation{Department of Physics, Ohio State University, Columbus, Ohio 43210, USA}
\affiliation{Department of Astronomy, Ohio State University, Columbus, Ohio 43210, USA}
\author{Annika H. G. Peter}

\affiliation{Center for Cosmology and AstroParticle Physics (CCAPP), Ohio State University, Columbus, Ohio 43210, USA}
\affiliation{Department of Physics, Ohio State University, Columbus, Ohio 43210, USA}
\affiliation{Department of Astronomy, Ohio State University, Columbus, Ohio 43210, USA}

\date{\daymonthyeardate\today}

\begin{abstract}
Critical probes of dark matter come from tests of its elastic scattering with nuclei. The results are typically assumed to be model independent, meaning that the form of the potential need not be specified and that the cross sections on different nuclear targets can be simply related to the cross section on nucleons.  For pointlike spin-independent scattering, the assumed scaling relation is $\sigma_{\chi A} \propto A^2 \mu_A^2 \sigma_{\chi N}\propto A^4 \sigma_{\chi N}$, where the $A^2$ comes from coherence and the $\mu_A^2\simeq A^2 m_N^2$ from kinematics for $m_\chi\gg m_A$.  Here we calculate where model independence ends, i.e., where the cross section becomes so large that it violates its defining assumptions.  We show that the assumed scaling relations generically fail for dark matter-nucleus cross sections $\sigma_{\chi A} \sim 10^{-32}-10^{-27}\;\text{cm}^2$, significantly below the geometric sizes of nuclei, and well within the regime probed by underground detectors. Last, we show on theoretical grounds, and in light of existing limits on light mediators, that pointlike dark matter cannot have $\sigma_{\chi N}\gtrsim10^{-25}\;\text{cm}^2$, above which many claimed constraints originate from cosmology and astrophysics. The most viable way to have such large cross sections is composite dark matter, which introduces significant additional model dependence through the choice of form factor. All prior limits on dark matter with cross sections $\sigma_{\chi N}>10^{-32}\;\text{cm}^2$ with $m_\chi\gtrsim 1\;\text{GeV}$ must therefore be reevaluated and reinterpreted. 
\end{abstract}

\maketitle


\section{Introduction}

The nature of dark matter is one of the most pressing problems in both fundamental physics and cosmology.  Decades of observations indicate that dark matter makes up the vast majority of matter in our Universe, yet increasingly advanced experiments have yet to determine its physical nature. Once discovered, the particle properties of dark matter will be a guidepost to physics beyond the Standard Model as well as to an improved understanding of galaxies and cosmic structure~\cite{Bertrev04,Pet12,bau12,Bertone:2016nfn, Plehn:2017fdg,Buckley:2017ijx,Baudis:2018bvr}. 

Progress depends on accurately assessing the regions of dark matter parameter space that remain viable. One of the best motivated dark-matter candidates is a single weakly interacting massive particle (WIMP). There are several ways to search for WIMPs:  first, through missing transverse momentum searches at colliders~\cite{Baltz:2006fm,Goodman:2010yf,Goodman:2010ku,Buchmueller:2014yoa,Abd15,ATLAS17,CMS18,Boveia:2018yeb}; second, through searches for WIMP self-annihilation products and decay~\cite{Delahaye:2007fr,Meade:2009iu,Hooper:2010mq,Steigman:2012nb,Albuquerque:2013xna,Bulbul:2014sua,Jeltema:2014qfa,Ackermann:2015zua,Leane:2018kjk, Abdalla:2018mve,Queiroz:2019acr,Cholis:2019ejx,Smirnov:2019ngs}; third, by energy transfer through elastic scattering with nuclei and electrons. Laboratory direct-detection experiments~\cite{Armengaud:2016cvl,  damic_result,Akerib:2016vxi,Amole:2017dex,Akerib:2017kat,xenon_1t,pandax_result,Agnes:2018fwg,supercdms_result1,dd9,Abdelhameed:2019hmk} provide the tightest bounds on dark matter-nucleus elastic scattering cross sections, with other constraints provided by cosmology and astrophysics~\cite{Cyb02, Gor10,Pro13,Dvorkin:2013cea,Kouvaris:2014lpa,Ali-Haimoud:2015pwa, Gluscevic:2017ywp,Bod18,Xu:2018efh,Slatyer:2018aqg,Bhoonah:2018wmw,Cappiello:2018hsu,Bringmann:2018cvk,Ema:2018bih,Gluscevic:2019yal,Wadekar:2019xnf,Nadler:2019zrb,Alvey:2019zaa,Cappiello:2019qsw}. While there are no robust signals yet, progress is rapid.
%
\begin{figure}[!htp]

\includegraphics[width=\columnwidth]{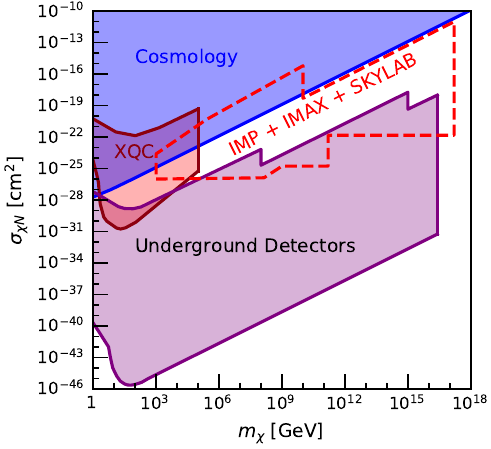} 
\caption{Claimed constraints on the spin-independent dark matter-nucleon cross section \cite{Wandelt:2000ad,Erickcek:2007jv,Kavanagh:2017cru,Mahdawi:2018euy,Nadler:2019zrb}. Those from cosmology directly probe scattering with protons, but all others here are based on scattering with nuclei, and thus require the use of  ``model-independent'' scaling relations.  Below, we show that assumptions used to derive these results are invalid over most of the plane.}
\label{fig:prexist_limit_simpl}
\end{figure}
%

For these searches, the two most common benchmarks for the performance of dark matter detection experiments are the dark matter self-annihilation cross section and the spin-independent dark matter-nucleon elastic scattering cross section, the simplest case (for more general treatments, see, e.g., Refs.~\cite{Fitzpatrick:2012ix,DelNobile:2013sia,Anand:2013yka,Peter:2013aha,Hoferichter:2015ipa,Gluscevic:2015sqa,Hoferichter:2018acd}). These benchmarks allow constraints set by different experiments to be scaled to each other. Here, we focus on spin-independent elastic dark matter-nucleus scattering for dark matter with $m_{\chi}\gtrsim 1$ GeV. For generality, we do not require that dark matter be a thermal relic.

Most direct-detection searches focus on pushing sensitivity to small cross sections, but it is also important to consider constraints on large cross sections \cite{Starkman:1990nj,Wandelt:2000ad,Albuquerque:2003ei,Erickcek:2007jv,Mack:2007xj,Albuquerque:2010bt,Mahdawi:2017cxz,Mahdawi:2017utm,Kavanagh:2017cru,Hooper:2018bfw,Emken:2018run,Bramante:2018qbc,Mahdawi:2018euy,Bhoonah:2018gjb}. Direct-detection experiments are typically located beneath the atmosphere, rock, and detector shielding, such that dark matter with too large of a cross section loses too much energy above the detector. Energy loss in the detector overburden may open a window where strongly interacting dark matter is allowed~\cite{Starkman:1990nj}.  

Figure~\ref{fig:prexist_limit_simpl} summarizes prior claimed constraints. The ``IMP+IMAX+SKYLAB'' region is based on atmospheric and space-based detectors and is dashed because the results are commonly cited but are not based on detailed analyses in peer-reviewed papers \cite{Wandelt:2000ad}. The X-ray Quantum Calorimeter (XQC) experiment is rocket based \cite{McCammon:2002gb}. There are several similar proposed XQC regions \cite{Wandelt:2000ad,Erickcek:2007jv,Mahdawi:2018euy}; we adopt that of  Ref.~\cite{Mahdawi:2018euy}. The ``Underground Detectors'' region is taken directly from the summary plot in Ref.~\cite{Kavanagh:2017cru}. For the ``Cosmology'' region, we plot the strongest constraint that depends only on dark matter-proton scattering \cite{Nadler:2019zrb} (including helium would make the constraints somewhat stronger \cite{Xu:2018efh}). The details of which constraints are plotted do not affect our conclusions. 

Direct-detection searches for spin-independent interactions benefit from an essentially model-independent $A^2$ coherent enhancement, as well as a kinematic factor of $\mu_A^2$, such that $\sigma_{\chi A}$ is related to the dark matter-nucleon elastic scattering cross section $\sigma_{\chi N}$ by $\sigma_{\chi A}\propto A^2\mu_A^2\sigma_{\chi N}$. For $m_\chi\gg m_A$, the dark matter-nucleus reduced mass $\mu_A\simeq A m_N$, such that the scaling becomes $\sigma_{\chi A}\propto A^4\sigma_{\chi N}$.  This straightforward scaling allows constraints on dark matter-nucleus scattering to be related to each other and to the cross section on nucleons.  This scaling is model independent in the sense that it is independent of the detailed shape of the potential. In Fig.~\ref{fig:prexist_limit_simpl}, all constraints except the one labeled ``Cosmology'' deal with nuclear targets with $A>1$, and hence assume this scaling relation.

How large of cross sections are allowed before the defining assumptions are violated?  Here we systematically calculate the theoretical upper limits on dark matter-nucleon cross sections.  We show that most of the parameter space of Fig.~\ref{fig:prexist_limit_simpl} is beyond the point where the simple scaling relations above are valid, or where pointlike dark matter is even allowed.  Our results are based first on generic considerations of theoretically allowed cross sections for short-range interactions with nuclei, and second on classes of models where we consider light mediators as a mechanism to obtain large cross sections. As far as we are aware, this is the first systematic exploration of these issues for dark matter-nucleus scattering (for related considerations in strongly self-interacting dark matter sectors, see, e.g., Ref.~\cite{Tulin:2013teo}). Our results will require the reinterpretation of a large and varied body of work, e.g. Refs.~\cite{Goodman:1984dc,PhysRevLett.61.510,Rich:1987st,Starkman:1990nj,McGuire:1994pq,Bernabei:1999ui,Derbin:1999uh,Wandelt:2000ad,Chen:2002yh,Albuquerque:2003ei,Zaharijas:2004jv,Mack:2007xj,Erickcek:2007jv, Albuquerque:2010bt,Dvorkin:2013cea,Jacobs:2014yca,Mahdawi:2017cxz,Mahdawi:2017utm,Kavanagh:2017cru,Gluscevic:2017ywp, Hooper:2018bfw,Emken:2018run,Xu:2018efh,Bramante:2018qbc,Slatyer:2018aqg,Mahdawi:2018euy,Bhoonah:2018gjb,Bramante:2018tos,Janish:2019nkk}.

In Sec.~\ref{sec:scattering_theory}, we review the nonrelativistic scattering theory used to obtain the model-independent scaling relations. In Sec.~\ref{sec:contact},  we examine the various ways that scaling relations can break down for contact interactions. In Sec.~\ref{sec:light_mediator}, we examine the possibility of achieving a larger cross section with a light mediator in light of present constraints on light mediators. In Sec.~\ref{sec:composite_dm}, we briefly discuss the possibility that dark matter itself could have a nonzero physical extent. In Sec.~\ref{sec:implications}, we discuss the implications for existing constraints and future experiments. Finally, we summarize our results and the outlook for future work in Sec.~\ref{sec:disc_conc}. 


\section{Dark Matter Scattering Theory}\label{sec:scattering_theory}
We briefly review the basic nonrelativistic scattering theory required to derive the model-independent scaling relation for the spin-independent elastic scattering cross section. We also discuss how some of the key assumptions may break down. Throughout, we set $\hbar=c=1$.

\subsection{Overview of basic assumptions}\label{ssec:basic_assumptions}

\begin{enumerate}
    \item {\bf Single particle: } Dark matter is primarily a single unknown particle. The number density of dark matter is then determined only by its mass and the local dark matter density. 
    \item {\bf Pointlike:} Dark matter is a pointlike particle with no excitation spectrum. 
    \item {\bf Electrically neutral:} It is typically assumed that dark matter is electrically neutral. Millicharged dark matter has different dynamics and is too strongly constrained to produce large cross sections. 
    \item {\bf Equal coupling to all nucleons:} For simplicity, we assume that dark matter has equal coupling to both protons and neutrons, although this assumption is not essential to any of our conclusions.
    \item {\bf Local:} The interaction is assumed to be local, $\left<{\bf x'}\left|\widehat{V}\right|{\bf x}\right>=V({\bf x'})\delta^3({\bf x'}-{\bf x})$.
    \item {\bf Energy-independent potential:} The potential for the interaction is assumed to be energy independent, such that the cross section for the interaction is also energy independent up to a form factor. For a spin-independent interaction, the potential must also be independent of the incident angular momentum $l$. 
    \item {\bf Elastic:} For laboratory experiments, dark matter-nucleus scattering is assumed to occur at typical Milky Way virial velocities, $v\sim 10^{-3}\,c$.  Typical recoil energies of $\mathcal{O}(1\;\text{keV})$ are not sufficient to produce Standard-Model particles, or to excite internal degrees of freedom of nuclei. Therefore, elastic scattering is the dominant interaction channel. In any case, all physical scattering processes have at least some elastic component \cite{joachain75_collision}.
    \item {\bf Coherence:} Closely related to the assumption of purely elastic scattering is the assumption of coherence. For coherence to hold, it must be a good approximation to treat the dark matter as interacting with the nucleus as a whole, rather than with individual nucleons. Coherence is typically a good approximation provided the momentum transfer $q$ is insufficient to excite internal degrees of freedom in a nucleus, which is true provided $1/q$ is large compared to the characteristic nuclear radius $r_A$, $q r_A\ll1$ \cite{bransden1983physics}. The breakdown of coherence can be parametrized by including a momentum-dependent form factor in the differential cross section. 
    \item {\bf No bulk effects:} The scattering should be well approximated as being with a single nucleus, such that initial and final state effects in the bulk medium can be ignored. This approximation is good as long as the characteristic momentum transfer $q$ is large compared to the characteristic interatomic spacing, which is typical.
 
\end{enumerate}
The rest of this paper deals with the failure of the following additional assumptions:
\begin{enumerate}[resume]
    \item {\bf S-wave scattering} For s-wave ($l=0$) scattering, the scattering is isotropic in the center-of-momentum frame. As shown in  Sec.~\ref{sssec:partial_waves}, assuming $l=0$ is required to derive the model-independent $A^2\mu_A^2$ scaling relation. However, real interactions may deviate significantly from isotropic scattering, and we do not require $l=0$ in this analysis. 
    \item {\bf Weak Interaction:} For $A^2\mu_A^2$ scaling to hold, the interaction must be weak enough for the Born approximation to hold. We discuss this assumption in Sec.~\ref{sssec:born_approx}.

\end{enumerate}


\subsection{Basic scattering theory}
Here we provide a brief review of the scattering theory formalism \cite{newton1964complex,regge65_scattering,joachain75_collision,goldberger1975collision,bransden1983physics,sakurai,khare2012introduction,burke_joachain} used in later sections.

To be detectable, dark matter must have some kind of interaction with ordinary matter in a detector, written here as a potential $V({\bf r})$. We specialize to spin-independent interactions, and restrict our analysis to spherically symmetric potentials $V({\bf r}) = V(r)$ that fall off faster than $r^{-1}$ as $r\rightarrow\infty$. In the center-of-momentum frame, the time-independent Schrödinger equation giving the evolution of a nonrelativistic two-particle system with wave function $\psi({\bf r})$ and reduced mass $\mu$ is given as
\begin{equation}\label{tise}
\left(-\frac{1}{2\mu}\nabla^2_{\bf r}+V({\bf r})\right)\psi({\bf r}) = E\psi({\bf r}).
\end{equation}
As shown in  Appendix~\ref{app:lippmann_schwinger},  far from the potential the solution of Eq.~\eqref{tise} may be written as
\begin{align}
\psi({\bf r})\xrightarrow[]{r\rightarrow\infty}&\psi_0({\bf r})-\frac{\mu e^{ikr}}{2\pi r} \int V({\bf r'})\psi({\bf r'})e^{-i{\bf k}_f\cdot{\bf r'}}d^3{\bf r'} \nonumber\\
=&\psi_0({\bf r})+(2\pi)^{-3/2}\frac{e^{ikr}}{r}f\left({\bf k}_i,{\bf k}_f\right),\label{scattering_amp_lipp}
\end{align}
where $f\left({\bf k}_i,{\bf k}_f\right)=f(k,\theta)$ is the scattering amplitude, $\psi_0({\bf r})\equiv(2\pi)^{-3/2}e^{i {\bf k}_i\cdot{\bf r}}$, and ${\bf k}_i\equiv k{\bf \hat{z}}$ and ${\bf k}_f$ are the initial and final dark matter momenta, respectively. From the scattering amplitude, we obtain the differential cross section
\begin{equation}\label{diff_scattering}
\frac{d\sigma}{d\Omega} =|f(k,\theta)|^2 
\end{equation}
and the total elastic scattering cross section:
\begin{equation}\label{tot_scattering}
\sigma_{\chi A}=\int \frac{d\sigma}{d\Omega} d\Omega.
\end{equation}

If the scattering is isotropic, $f(k,\theta)=f(k)$, Eq.~\eqref{tot_scattering} is proportional to the rate of detectable scattering events in a detector. However, to be detectable, a collision must deposit sufficient energy into the detector. If the scattering angle is peaked close to $\theta=0$, very little momentum is transferred, and hence insufficient energy is deposited in the detector. Therefore, it is sometimes more useful to weight the integral in Eq.~\eqref{tot_scattering} by the momentum transfer to obtain the momentum-transfer cross section,
\begin{equation}\label{mt_scattering}
\sigma^{\text{mt}}_{\chi A}=\int \frac{d\sigma}{d\Omega}\left(1-\cos\theta\right) d\Omega.
\end{equation}

For isotropic scattering, $\sigma^{\text{mt}}_{\chi A}=\sigma_{\chi A}$. For a potential with characteristic radius $r_A$, isotropic scattering is generically a good approximation at low energies, $k r_A\ll 1$, as discussed further in  Sec.~\ref{sssec:partial_waves}. Forward scattering is a major concern for light mediators (Sec.~\ref{sec:light_mediator}); in the Coulomb scattering limit where the mediator mass $m_\phi\rightarrow0$, then $\sigma_{\chi A}\rightarrow\infty$, while $\sigma^{\text{mt}}_{\chi A}$ remains finite. 

\subsection{Derivation of model-independent scaling}
Now we discuss approximation methods for $f(k,\theta)$. The two approaches we consider here are the Born approximation and the partial wave expansion. Both approaches allow us to derive the $\sigma_{\chi A}=A^2\left(\mu_A^2/\mu_N^2\right)\sigma_{\chi N}$ scaling with nuclear mass number $A$. The reduced masses are defined as $\mu_A\equiv m_A m_\chi/(m_A+m_\chi)$, $\mu_N\equiv m_N m_\chi/(m_N+m_\chi)$, where the mass of a nucleus with mass number $A$ is related to the mass of a single nucleon $m_N$ using $m_A= Am_N$. We begin with the Born approximation because it is the simple and familiar derivation. Because the partial wave expansion is valid even when the Born approximation fails, it allows us to more concretely show the behavior at large scattering cross sections. 

\subsubsection{Born approximation}\label{sssec:born_approx}
Inspecting Eq.~\eqref{scattering_amp_lipp}, a natural first approach to obtaining $f(k,\theta)$ is to solve for $\psi({\bf r})$ by iteration, which is the Born approximation, as demonstrated in Appendix~\ref{app:born_der}. The first Born approximation to $f(k,\theta)$ is simply the Fourier transform of the potential:
\begin{equation}\label{first_born_fourier_body}
f^{(1)}\left(k,\theta\right)=f^{(1)}\left(q\right)=-\frac{2\mu_A}{q}\int_0^\infty V(r')\sin(qr')r' dr', 
\end{equation}
where $q=|{\bf q}|=2k\sin\theta/2$ is the momentum transfer. 

Now, assume that the potential has some maximum radius $r_A$, and we have low energy scattering, $k r_A\ll 1$. Then we can approximate $\sin(qr')\approx q r'$ and integrate only up to the maximum radius $r_A$:
\begin{equation}\label{potential_ind_approx}
f^{(1)}\left(k,\theta\right)\approx-2\mu_A\int_0^{r_A} V(r')r'^2 dr'.
\end{equation}

Equation~\eqref{potential_ind_approx} is a remarkable result. If the required approximations are valid, $f^{(1)}\left(k,\theta\right)$ depends only on the volume integral of the potential; it contains no information at all about the shape. Provided the volume integral of the potential is proportional to the nuclear mass number $A$, we have the scaling
\begin{equation}\label{model_ind_scaling_amp}
f^{(1)}\left(k,\theta\right)\propto A\mu_A.
\end{equation}
Plugging into Eq.~\eqref{tot_scattering}:
\begin{equation}\label{model_ind_scaling_prop}
\sigma^{(1)}_{\chi A}\propto A^2\mu_A^2,
\end{equation}
which can be recast more precisely in terms of the dark matter-nucleon reduced mass $\mu_N$ and scattering cross section $\sigma_{\chi N}^{(1)}$:
\begin{equation}\label{model_ind_scaling_tot}
\sigma^{(1)}_{\chi A}= A^2\frac{\mu_A^2}{\mu_N^2}\sigma_{\chi N}^{(1)}.
\end{equation}

 Eq.~\eqref{model_ind_scaling_tot} is the famous model-independent scaling relation for the spin-independent elastic scattering cross section. Provided the potential falls off faster than $1/r$, this scaling relation is generally a good approximation at sufficiently low energies, so long as the first Born approximation reasonably approximates $f(k,\theta)$. However, we must examine when the first Born approximation fails. 

We discuss the validity of the Born approximation in Appendix~\ref{app:born_der}. A useful condition for the validity of the first Born approximation is \cite{khare2012introduction}
\begin{equation}\label{final_cond_born_use}
\frac{\mu_A}{k}\left|\int_0^\infty V(r')\left(e^{2ikr'}-1\right)dr'\right|\ll 1.
\end{equation}

We can simplify Eq.~\eqref{final_cond_born_use} using our assumption of a maximum range $r_A$ and $k r_A\ll 1$:
\begin{equation}\label{born_cond_approx}
2\mu_A\left|\int_0^{r_A} V(r')r'dr'\right|\ll 1.
\end{equation}
Equation~\eqref{born_cond_approx} is equivalent to the statement that the potential is much too weak to form a bound state even if $V(r)$ was purely attractive \cite{levinson1949uniqueness}. While Eq.~\eqref{potential_ind_approx} is a volume integral, Eq.~\eqref{born_cond_approx} is an area integral of the potential. Therefore, the first Born approximation is valid when some potential-weighted effective area is small. The effective area in question is, in fact, the elastic scattering cross section, as shown for a contact interaction in Sec.~\ref{sec:contact}. 

\subsubsection{Partial wave expansion}\label{sssec:partial_waves}

To investigate what happens when the Born approximation fails, the first step is to expand the scattered wave function in terms of Legendre polynomials and calculate the phase shift of each contribution. The phase shifts may be found by numerically integrating the Schrödinger equation, as described in Appendix~\ref{app:partial_wave}. The elastic scattering cross section may be written in terms of the phase shifts $\delta_l(k)$: 

\begin{equation}\label{partial_wave_cross_sec}
\sigma_{\chi A}=\frac{4\pi}{k^2}\sum_{l=0}^\infty{(2l+1)\sin^2\left(\delta_l\right)}.
\end{equation}

The momentum-transfer cross section in Eq.~\eqref{mt_scattering} may also be written in terms of partial wave phase shifts:

\begin{equation}\label{partial_mt_cross_section}
\sigma^{\text{mt}}_{\chi A}=\frac{4\pi}{k^2}\sum_{l}(l+1)\sin^2(\delta_{l+1}-\delta_l). 
\end{equation}
The mathematical decomposition in terms of partial waves is valid even beyond interactions that can be described in nonrelativistic potential scattering theory. However, when the number of partial waves becomes too large, it may be impractical to compute the phase shifts individually, and semiclassical approximations become useful \cite{joachain75_collision,sakurai}. 

Physically, the sum over partial waves in Eq.~\eqref{partial_wave_cross_sec} is equivalent to the classical operation of averaging over all possible impact factors $b=\sqrt{l(l+1)}/k$ \cite{burke_joachain, joachain75_collision}. Classically, for a potential with maximum range $r_A$, there would be no collisions for $b>r_A$. Therefore, a useful approximate upper limit on the highest partial wave that can meaningfully contribute to the sum in nonrelativistic quantum scattering is $l_{\text{max}}\approx k r_A$, and contributions from higher $l>l_{\text{max}}$ fall off quickly \cite{joachain75_collision}. Our derivation of the model-independent scaling of Eq.~\eqref{model_ind_scaling_tot} in the Born approximation assumes $k r_A\ll 1$, which is equivalent to saying only the $l=0$ (s-wave) term contributes a nonvanishing phase shift.

Using the same iterative procedure as for the Born approximation, we can obtain the model-independent form of the s-wave phase shift \cite{burke_joachain}:
\begin{equation}\label{swave_model_ind}
\delta_0(k)\approx -2\mu_A k\int_0^{r_A} V(r')r'^2 dr',
\end{equation}
where the required approximation is $\delta_0(k)\ll 1$. Plugging Eq.~\eqref{swave_model_ind} into Eq.~\eqref{partial_wave_cross_sec} and again using $\delta_0(k)\ll 1$, we obtain precisely the same expression for the scattering amplitude we obtained for the Born approximation in Eq.~\eqref{potential_ind_approx}, as expected. Again, the requirement $\delta_0(k)\ll 1$ places an upper bound on the maximum $\sigma_{\chi A}$ where the relation can apply. 

If the cross section were instead the maximum allowed by unitarity, $\delta_0(k)=\pi/2$, we would obtain
\begin{equation}\label{swave_max_cross}
\sigma_{\chi A}=\frac{4\pi}{k^2}, 
\end{equation}
which \emph{decreases} as $1/A^2$ with increasing $A$, assuming $k\propto A$, rather than increasing as $A^4$.

Higher partial waves necessarily scale differently with $k$ \cite{joachain75_collision}, $\delta_l(k)\propto k^{2l+1}$ for $k r_A\ll 1$. Higher $\delta_l(k)$ also contain information about the shape of the potential. Therefore, we do not expect any special model-independent scaling when higher partial waves contribute. 


\section{Contact interactions}\label{sec:contact}
In this section, we consider the limits on cross sections that can be obtained through a contact interaction, and how the scaling relations break down. A contact interaction is useful as an illustrative case because we do not need to consider the specific mechanism that produces the interaction.

\subsection{Contact interaction with Born approximation}\label{ssec:contact_born}

As a simple case, we consider a contact interaction with a nucleus, as could be produced by a heavy mediator. We roughly approximate the nuclear charge density as having a top-hat shape with radius $r_A$:
\begin{equation}\label{potential_well}
V(r)=
\begin{cases}
V_0 &r<r_A\\
0 &\text{otherwise}.
\end{cases}
\end{equation}
We assume the maximum charge density is roughly independent of atomic mass number $A$, such that $r_A\approx A^{1/3}r_N$, where $r_N\simeq 1.2\;\text{fm}$.

We use this toy model with a sharp cutoff because both the Born approximation and the partial wave phase shift $\delta_l(k)$ can be found analytically. The effect of using a more realistic charge distribution is discussed in Sec.~\ref{sssec:realistic_charge}.

Fourier transforming Eq.~\eqref{potential_well} using Eq.~\eqref{first_born_fourier_body} gives
\begin{equation}
f^{(1)}(q)=\frac{2\mu_A V_0}{q^3}\left[qr_A\cos(q r_A)-\sin(q r_A)\right].
\end{equation}

The total elastic scattering cross section in the first Born approximation is then:
\begin{multline}\label{tophat_1born}
\sigma^{(1)}_{\chi A}=\frac{\pi\mu_A^2 V_0^2}{16k^6}\big[4kr_A\sin(4kr_A)+\cos(4kr_A)\\
+32k^4r_A^4-8k^2r_A^2-1\big].
\end{multline}
In the limit $kr_A\ll 1$, Eq.~\eqref{tophat_1born} becomes
\begin{equation}\label{tophat_1born_lowk}
\sigma^{(1)}_{\chi A}\approx\frac{16\pi}{9}\mu_A^2r_A^6V_0^2.
\end{equation}
For scattering with a nucleon, Eq.~\eqref{tophat_1born_lowk} would become $\sigma^{(1)}_{\chi N}\approx\frac{16\pi}{9}\mu_N^2r_N^6V_0^2$. Substituting $r_A\approx A^{1/3}r_N$, we recover the required scaling relation of Eq.~\eqref{model_ind_scaling_tot}:
\begin{equation}\label{tophat_scaling}
\sigma^{(1)}_{\chi A}\approx A^2\frac{\mu_A^2}{\mu_N^2}\sigma^{(1)}_{\chi N}.
\end{equation}

In the $k r_A\ll 1$ limit, the condition for validity of the first Born approximation in Eq.~\eqref{born_cond_approx} is simply
\begin{equation}\label{tophat_1born_lowk_cond}
\mu_A r_A^2 V_0\ll 1.
\end{equation}
Comparing to Eq.~\eqref{tophat_1born_lowk}, we can rewrite the condition Eq.~\eqref{tophat_1born_lowk_cond} as:
\begin{equation}\label{tophat_1born_lowk_crosscond}
\sigma^{(1)}_{\chi A}\ll \frac{16}{9}\pi r_A^2.
\end{equation}
Equation~\eqref{tophat_1born_lowk_crosscond} has a simple physical interpretation. The first Born approximation is only applicable for elastic scattering cross sections much smaller than the geometric cross section of the nucleus. Using $r_N\approx 1.2\;\text{fm}$ in Eq.~\eqref{tophat_1born_lowk_crosscond}, the Born approximation result only applies for $\sigma^{(1)}_{\chi N}\ll 10^{-25}\;\text{cm}^2$. 

Going to higher orders in the Born approximation does not unlock cross sections significantly exceeding the geometric limit of the potential. For $\sigma^{(1)}_{\chi A}>\frac{16}{9}\pi r_A^2$, the Born series is not even guaranteed to converge for all energies \cite{joachain75_collision}. 

However, it may still be possible to obtain a meaningful cross section in regimes where the Born approximation fails using partial wave analysis. We explore this technique below.

\subsection{Contact interaction with partial waves}\label{ssec:tophat_swave}

For $r<r_A$ in Eq.~\eqref{potential_well}, the radial wave function decomposed in partial waves has an analytic solution in terms of partial waves, $u_l(r)=C_l r j_l(k' r)$, where $k'\equiv\left(k^2-2\mu_A V_0\right)^{1/2}$ could be either pure real or pure imaginary. First, we consider the s-wave cross section, with $l=0$. Expanding in the limit where $V_0$ and $k$ are small, $(kr_A)^2\ll 1$, $|V_0|\ll 1/(2\mu_A r_A^2)$, we recover (see Appendix~\ref{app:partial_wave})

\begin{equation}\label{delta_l_tophat_order2}
\delta_0(k)\approx-\frac{2\mu_A k r_A^3V_0}{3}+\frac{8\mu_A^2kr_A^5V_0^2}{15}+\mathcal{O}\left(|V_0|^3\right).
\end{equation}
The corresponding s-wave cross section is
\begin{equation}\label{s_wave_sigma_order2_tophat}
\sigma^{l=0}_{\chi A}\approx\frac{16\pi}{9} \mu_A^2 r_A^6 V_0^2-\frac{128\pi}{45}\mu_A^3r_A^8V_0^3+\mathcal{O}\left(|V_0|^4\right),
\end{equation}
which is identical to Eq.~\eqref{tophat_1born_lowk} to lowest order in $|V_0|$, as anticipated in Sec.~\ref{sssec:partial_waves}. 

Now we can see how the model-independent $A$ scaling fails as the coupling strength gets stronger. The second-order term in Eq.~\eqref{s_wave_sigma_order2_tophat} scales $\propto\mu_A^3/\mu_N^3 A^{8/3}$, and either reduces the cross section for a repulsive potential $V_0>0$ or increases it for an attractive one. For $V_0$ large enough for the second-order and higher corrections to matter, there is, therefore, no model-independent scaling with $A$ because $\sigma^{l=0}_{\chi A}$ depends on the details of the potential. The breakdown of the scaling as a function of $A$ is shown in Fig.~\ref{fig:scale_saturation}. Once details of the potential begin to matter, corrections from a more realistic charge distribution would also become important, as discussed in Sec.~\ref{sssec:realistic_charge}.

To illustrate further, if we instead considered the strong coupling limit, $|V_0|\gg1/{2\mu_A r_A^2}$, for $V_0>0$ we would obtain: 

\begin{equation}\label{large_V0_delta}
\delta_0(k)=-k r_A,
\end{equation}

and
\begin{align}\label{sigma_s_wave_large_V0}
\sigma^{l=0}_{\chi A}&=\frac{4\pi}{k^2}\sin^2\left(-k r_A\right)\\
&=4\pi r_A^2,\label{tophat_geo_hard_sphere}
\end{align}
so the repulsive cross section completely saturates at 4 times the geometric cross section. The saturation is plotted as a function of $|V_0|$ in Fig.~\ref{fig:v0_scale_figure}. Physically, the well has simply become an impenetrable hard sphere with a fixed radius. Therefore, we have obtained a physical maximum cross section for the repulsive contact interaction at small $k$ which is only $9/4$ times larger than the maximum we obtained using the first Born approximation.

\subsubsection{Higher partial waves}\label{sssec:higher_partial}
While the s-wave cross section is limited to the geometric cross section, it is natural to wonder if the contributions from higher partial waves could allow a larger total cross section. For $k r_A\gg1$, we can approximate semiclassically $l_\text{max}\approx k r_A$, as described in Sec.~\ref{sssec:partial_waves}. Of course, in the quantum case, it is possible for higher partial waves to contribute, but their contributions fall off rapidly for $l>l_\text{max}$. Assuming a sharp cutoff is adequate for deriving an approximate maximum physically achievable cross section. The maximum possible cross section is achieved by saturating partial wave unitarity, i.e., taking $\delta_{l\leq l_\text{max}}(k)=\pi/2$:
\begin{align}\label{partial_wave_max_tophat}
\sigma_{\chi A}&=\frac{4\pi}{k^2}\sum_{l=0}^{l_\text{max}}(2l+1)\\
&=\frac{4\pi}{k^2}(1+l_\text{max})^2\\
&\approx 4\pi r_A^2.\label{partial_wave_max_tophat_fin}
\end{align}
Now, we see that the saturation at approximately the geometric cross section, found for $k r_A\ll 1$ in Eq.~\eqref{tophat_geo_hard_sphere}, also holds for $k r_A\gg1$.

In fact, for a very strong repulsive contact interaction, the phase shifts for $l\leq l_\text{max}$ approach $\delta_{l\leq l_\text{max}}(k)\approx l\pi/2-kr_A$ \cite{joachain75_collision}, such that $\sigma_{\chi A}\approx 2\pi r_A^2$. Therefore, a repulsive hard sphere almost saturates the unitarity limit of Eq.~\eqref{partial_wave_max_tophat_fin}. 

Including higher partial waves is therefore not a useful way of increasing the cross section, because the potential remains limited by the characteristic radius $r_A$.

Figure~\ref{fig:sigma_sigma} shows the breakdown of the $A^4$ scaling for several example nuclei, fully taking into account contributions from higher partial waves. Direct-detection constraints for underground detectors are  affected by the breakdown of scaling at the $\mathcal{O}(1)$ level for $\sigma_{\chi N}\simeq10^{-32}\;\text{cm}^2$.

\subsection{Attractive resonances}\label{ssec:attract_res}

A final possible approach would be to saturate unitarity at  $\delta_{l}=\pi/2$ while $k r_A \ll 1$, such that
\begin{equation}\label{swave_max}
\sigma_{\chi A}^{\text{max}}=\frac{4\pi}{k^2}(2l+1)
\end{equation}
can become large. The limit $\delta_{l}=\pi/2$  at $k r_A\lesssim 1$ can be achieved through resonances, which occur when an attractive potential becomes strong enough to support a bound state.

In reality, the resonant scattering cross section would achieve large values only for a narrow range of $k$ relative to the incident dark matter velocity distribution \cite{breit_wigner,joachain75_collision}. Because $k=\mu_A v$, the resonances are generically at different incident dark matter velocities for different elements, which guarantees that there are not any useful model-independent scaling relations relating the observed cross sections for strongly attractive potentials between different target materials. 

\begin{figure}[t]
\includegraphics[width=\columnwidth]{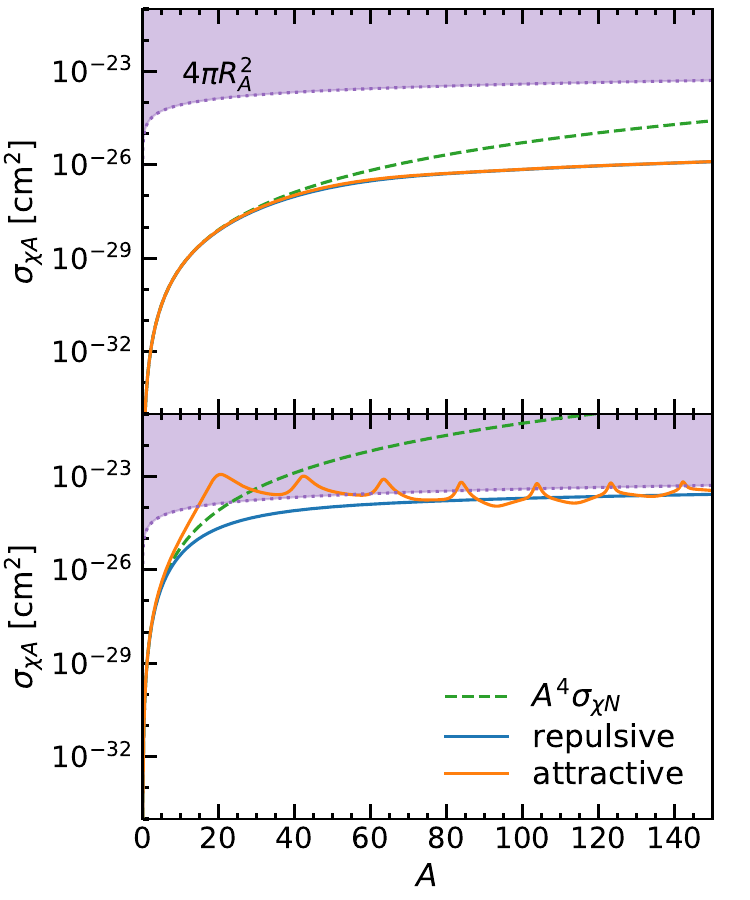}
\caption{{\bf Top:} Scaling with $A$ for the contact interaction in Sec.~\ref{sec:contact} with $|V_0|=1.18\times10^{-5}\;\text{GeV}$, computed using $k=0.005 A\;\text{fm}^{-1}$, $R_A=1.2 A^{1/3}\;\text{fm}$. We include partial waves up to $l_\text{max}=8$, which is sufficient to converge $\sigma_{\chi A}$ to $\sim 10^{-16}$ precision. Attractive and repulsive interactions scale similarly, although the scaling deviates from $A^4$ at high $A$ due to form-factor suppression, accounted for here by including the contributions from higher partial waves. 
\\
{\bf Bottom:} Same as above, but with $|V_0|=1.18\times10^{-3}\;\text{GeV}$, which corresponds to the `scaling relations unreliable for $A>12$' line in Fig.~\ref{fig:scattering_scaling}. Repulsive and attractive interactions no longer scale the same way, and both saturate close to $4\pi R_A^2$. The attractive potential shows resonances with $A$, which are sensitive to the specific choice of potential. For cross sections approaching the geometric cross section, any scaling with $A$ is highly model dependent.}\label{fig:scale_saturation}
\end{figure}

\begin{figure}[t]
 \includegraphics[width=\columnwidth]{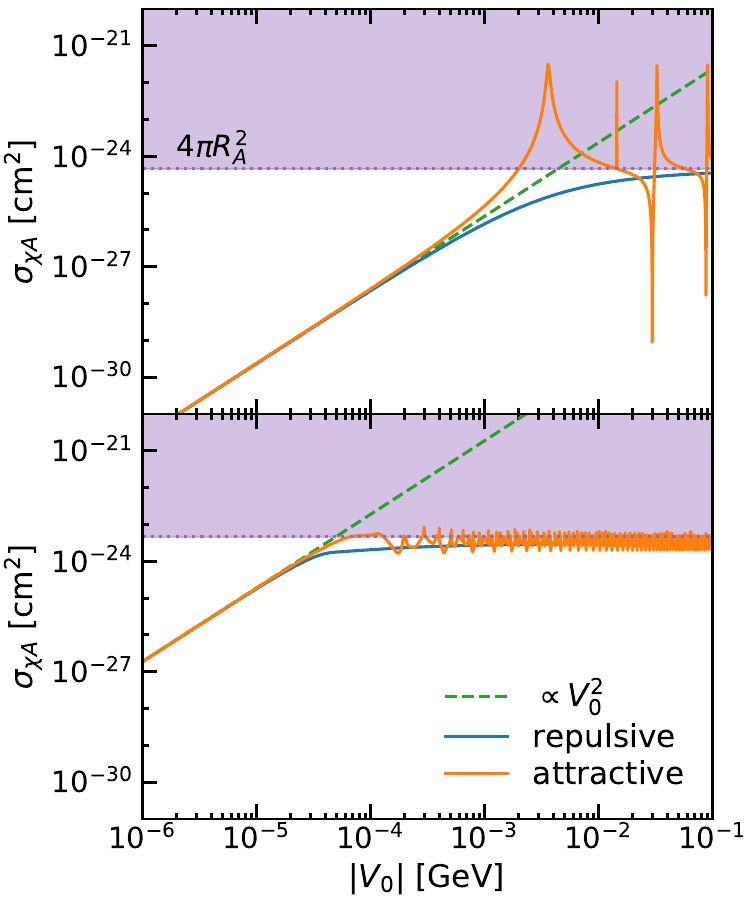}
 \caption{{\bf Top}: Scaling of cross section with $|V_0|$ for $A=4$ (helium), calculated using the contact interaction in Sec.~\ref{sec:contact}. The cross sections are computed using the analytic partial wave results. For attractive potentials, once $|V_0|$ becomes large enough to support quasibound states, resonances can increase the cross section by several orders of magnitude, but only in a narrow range.  \\
 {\bf Bottom}: Same as above, but with $A=131$ (xenon). A larger number of partial waves contribute due to the larger $k\propto A$. There are many resonances, but they are not large enough to meaningfully increase the cross section above the geometric limit. Additionally, the resonances are not at the same values of $|V_0|$, which prevents resonances from achieving a large cross section which scales predictably with $A$, as shown in Fig.~\ref{fig:scale_saturation}.}
\label{fig:v0_scale_figure}

\end{figure}

\begin{figure}[tp]
    \includegraphics[width=\columnwidth]{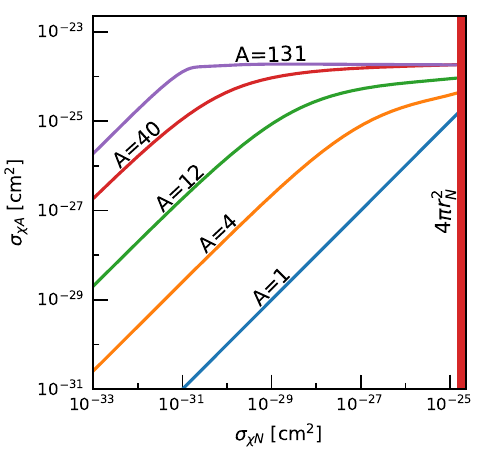} \\
    \caption{Scaling of the nuclear cross section with nucleon cross section for the repulsive contact interaction of Sec.~\ref{sec:contact} at fixed $k_N=0.005\;\text{fm}^{-1}$. The contact interaction cannot achieve nucleon cross sections larger than the geometric cross section, denoted by the vertical red line. The cross section visibly deviates from $A^4$ scaling at the $\mathcal{O}(1)$ level for heavy nuclei even for $\sigma_{\chi N}\simeq10^{-32}\;\text{cm}^2$, and by the time scaling fails at the $\mathcal{O}(1)$ level for $^4$He at $\sigma_{\chi N}\simeq4\times10^{-28}\;\text{cm}^2$, the cross sections for heavy nuclei have completely saturated. The scaling could break down in different ways in other models.}
    \label{fig:sigma_sigma}
\end{figure}

In Fig.~\ref{fig:scale_saturation}, we show the behavior as a function of $A$ for two different values of $V_0$. When resonances are possible, the scaling with $A$ need not be monotonic. The behavior is fairly complex for even the simple rectangular well of Eq.~\eqref{potential_well}. Realistic scaling is likely to be even more complicated, because the nuclear charge distribution changes as a function of $A$. Even two different nuclei with the same $A$ but different atomic numbers could have different charge distributions, and hence different resonant cross sections. Scaling with $A$ for strong attractive couplings is therefore highly model-dependent. 
In Fig.~\ref{fig:v0_scale_figure}, we show the saturation of the s-wave cross section as a function of the coupling strength, as well as the resonant behavior which occurs once the potential becomes strong enough to support a quasibound state. The resonances for $A=4$ are fairly narrow, as for a nucleon the scattering is still well approximated in the low-$k$ limit. However, for $A=131$, the low-$k$ limit is no longer a good approximation, and resonances are broadened to the point that they do not significantly increase the scattering cross section. Additionally, there are many more resonances, from multiple partial waves. Note we have not implemented any velocity dispersion for this plot; the spreading of resonances is entirely due to broadening of peaks and overlapping contributions from multiple partial waves at finite $k$. Applying a realistic dark matter velocity distribution would smooth the peaks. For heavy nuclei with multiple naturally occurring isotopes (e.g. xenon), averaging over a distribution of isotopes would smooth the peaks even further.

Overall, even if a carefully tuned resonance could achieve a large cross section for a single light nucleus, other nuclei would not necessarily have correspondingly large cross sections. Scaling relations between specific nuclear cross sections would also be highly model dependent, such that constraints from different types of nuclei would be difficult to compare because the full resonance structure would not be known. For example, using more realistic charge distributions, such as an exponential potential for $A=4$ and a Woods-Saxon potential for $A=131$ \cite{charge_dist_1987} would shift the positions of the resonances somewhat.

\subsubsection{More realistic charge distributions}\label{sssec:realistic_charge}
The rectangular barrier potential in Eq.~\eqref{potential_well} is a toy model. Realistic nuclear charge distributions have a smooth cutoff, and an exponential tail to larger radii \cite{charge_dist_1987}, as in a Woods-Saxon potential. Because there is not a sharp cutoff, allowing the interaction strength to be arbitrarily large would cause the potential to grow logarithmically with $|V_0|$. However, there are limits to how strong $|V_0|$ can be. For $|V_0|\gtrsim 10^{-1}\;\text{GeV}$, QCD corrections break the simple nonrelativistic contact interaction picture. For $|V_0|\gtrsim 2\;\text{GeV}$, the interaction may be strong enough to pull proton-antiproton pairs out of the vacuum.

We have verified by numerically computing partial wave amplitudes that, for $|V_0|\lesssim 10^{-1}\;\text{GeV}$, using a Woods-Saxon potential increases the maximum $\sigma_{\chi A}$ by a factor of $\lesssim 10$. By definition, an increase in the computed cross section due to a different potential can only appear for $|V_0|$ strong enough that the model-independent form of the cross section has already significantly broken down. Therefore, using a more realistic charge distribution cannot significantly change our conclusion that the cross section for a contact interaction cannot be much larger than the geometric cross section of the nucleus. 

Realistic potentials are also not perfectly spherically symmetric; however, the same basic picture of geometrical limitations still applies, and resonances are still possible.

\subsection{Beyond contact interactions}\label{ssec:beyond_contact}
We have established that a contact interaction with a nucleus cannot achieve cross sections much larger than the geometrical cross section of the nucleus. The case where large cross sections might be achieved, a strongly attractive potential, produces resonances that are sensitive to the detailed structure of the potential and is far too model dependent to possess any simple scaling relation relating the cross sections at different $A$. To circumvent these problems, we need an interaction with a larger characteristic range. One possible way to achieve a larger characteristic range is to insert a light mediator for the interaction, as discussed in Sec.~\ref{sec:light_mediator}. Another possibility is composite dark matter with an intrinsic radius, discussed in Sec.~\ref{sec:composite_dm}.


\section{Light Mediator}\label{sec:light_mediator}

A simple approach to achieving a larger characteristic radius is to insert a light mediator, of mass $m_\phi=1/r_\phi$, which generically results in a potential of the form
\begin{equation}\label{yukawa_potential}
V(r) = \frac{\lambda_A\lambda_\chi}{4\pi}\frac{e^{-r/r_\phi}}{r},
\end{equation}
where $\lambda_\chi$ and $\lambda_A=A\lambda_N$ are the coupling strengths of the particle $\phi$ to the dark matter and nucleus respectively. To achieve cross sections much larger than a nucleus, we should have $r_\phi\gg1\;\text{fm}$. The dark matter and target nucleus are distinguishable particles, so the Yukawa potential can be either attractive or repulsive. We assume the mediator is a scalar, although the general form of the potential would be similar for other light mediator candidates. The scattering amplitude of Eq.~\eqref{yukawa_potential} is easily calculated using Eq.~\eqref{first_born_fourier_body}:
\begin{equation}\label{yukawa_fourier}
f^{(1)}\left(q\right)=-\frac{\mu_A\lambda_\chi\lambda_A}{2\pi(q^2+1/r_\phi^2)},
\end{equation}
which gives the total elastic scattering cross section:
\begin{equation}\label{tot_cross_yukawa}
\sigma^{(1)}_{\chi A}=\frac{\mu_A^2\lambda_\chi^2\lambda_A^2r_\phi^4}{\pi(1+4k^2r_\phi^2)}.
\end{equation}
Because the characteristic radius is larger than the geometric radius of a nucleon, Eq.~\eqref{tot_cross_yukawa} can, in principle, achieve larger cross sections within the domain of validity of the Born approximation than a contact interaction could. Because $k\propto \mu_A$, the scaling with $A$ is now more complicated due to the $k^2 r_\phi^2$ term in the denominator. Assuming $m_\chi\gg m_A$ such that $\mu_A\approx A m_N$, we have two limits:

\begin{equation}\label{sigma_yukawa_limits}
\sigma^{(1)}_{\chi A}\approx \begin{cases}
A^4\sigma^{(1)}_{\chi N} & k_A r_\phi\ll 1\\
A^2\sigma^{(1)}_{\chi N} & k_A r_\phi\gg1.
\end{cases}  
\end{equation}

For direct detection, $k_N\sim0.005\;\text{fm}^{-1}$ is set by Milky Way halo velocities $v\simeq10^{-3}\,c$ and the mass of a single nucleon, $k_A\simeq A k_N$. For $^{131}$Xe, $k_A\approx0.7\;\text{fm}^{-1}$, such that $k_A r_\phi>1$ occurs for $r_\phi\gtrsim1.4\;\text{fm}$. Therefore, $A^4$ scaling is, at best, marginal for heavy nuclei for any $r_\phi$ that could conceivably produce a cross section $\sigma_{\chi A}\gtrsim 10^{-25}\;\text{cm}^2$. Additionally, as established for a general potential with a characteristic radius $r_\phi$ in Sec.~\ref{sssec:born_approx}, increasing the coupling strengths $\lambda_\chi$ or $\lambda_N$ at fixed $r_\phi$ causes the Born approximation, and therefore the $A^4$ scaling, to fail before cross sections larger than the geometric cross section are achieved. Therefore, for a light mediator, the $A^4$ scaling is only possible if $\sigma^{(1)}_{\chi A}\ll 10^{-25}\;\text{cm}^2$. The failure of $A^4$ scaling occurs without even considering the constraints on the existence of light mediators discussed in Sec.~\ref{ssec:yukawa_med_lim}. The $A^4$ scaling is preserved to somewhat larger cross sections than for the contact interaction shown in Fig.~\ref{fig:scale_saturation}.
\begin{figure}[tp]
    \includegraphics[width=\columnwidth]{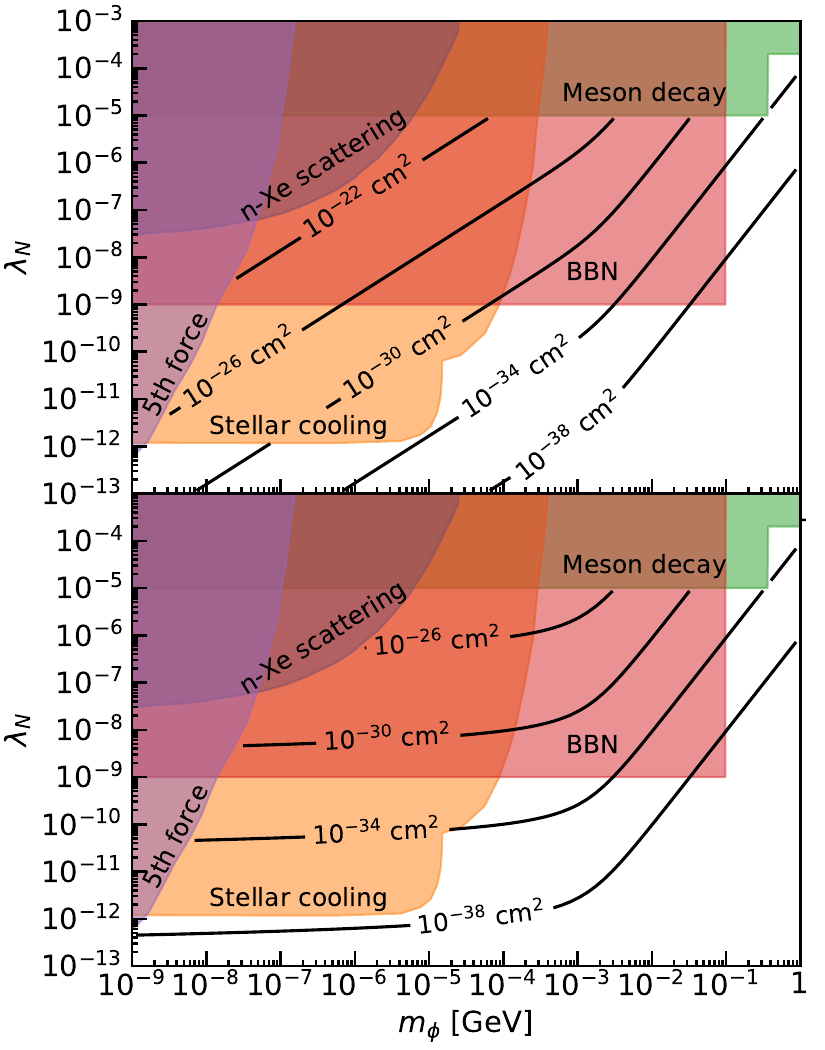}
 \caption{{\bf Top:} Elastic scattering cross-section contours as a function of mediator mass and coupling strength for the repulsive Yukawa potential in Eq.~\eqref{yukawa_potential}. We also show various constraints on the existence of such mediators from Ref.~\cite{light_mediator}. The largest cross sections achieved in unconstrained regions are $\sigma_{\chi N}\lesssim10^{-27}\;\text{cm}^2$. For $m_\phi<10^{-9}\;\text{GeV}$, fifth-force constraints become many orders of magnitude stronger and dominate other constraints \cite{Murata:2014nra}.\\
 {\bf Bottom:} Same as above, but for the momentum-transfer cross section. The largest cross sections achieved in unconstrained regions are $\sigma^{\text{mt}}_{\chi N}\lesssim10^{-32}\;\text{cm}^2$.}
    \label{fig:combine_cross}
\end{figure}
\subsection{Momentum-transfer cross section}
In fact, even the $A^2$ scaling is too optimistic for the detectable momentum transfer in a detector with high $A$. Inspection of Eq.~\eqref{yukawa_fourier} shows that for $k r_\phi\gg1$, the scattering becomes strongly peaked at $\theta=0$. Therefore, it is more useful to consider the momentum-transfer cross section, Eq.~\eqref{mt_scattering}.
Using the Born approximation, we can calculate Eq.~\eqref{mt_scattering} analytically for the Yukawa potential:
\begin{multline}\label{mt_yukawa}
\sigma^{\text{mt},(1)}_{\chi A}=\frac{\mu_A^2\lambda_\chi^2\lambda_A^2}{8\pi k^4(1+4k^2r_\phi^2)}\\
\times\big[(1+4 k^2r_\phi^2)\log(1+4 k^2r_\phi^2)
-4k^2r_\phi^2\big].
\end{multline}
 For $k r_\phi\ll 1$, Eq.~\eqref{mt_yukawa} simplifies to $\sigma^{\text{mt},(1)}_{\chi A}\approx\sigma^{(1)}_{\chi A}$ as expected for isotropic scattering. However, for $k r_\phi\gg1$, we have:
 \begin{equation}\label{mt_yukawa_highk}
 \sigma^{\text{mt},(1)}_{\chi A}\approx\frac{\mu_A^2\lambda_\chi^2\lambda_A^2}{8\pi k^4}\left(\log(4k^2r_\phi^2)-1\right).
 \end{equation}
Equation~\eqref{mt_yukawa_highk} grows only $\propto \log(A)$, such that, for a fixed total detector mass, the total energy deposited in the detector would be larger for nuclei with \emph{smaller} $A$. Direct detection experiments that focus on protons and other light nuclei, such as Refs.~\cite{Behnke:2016lsk,Amole:2017dex,Lippincott:2017yst,Collar:2018ydf}, may therefore be effective ways of constraining the landscape for model-dependent direct detection.

\subsection{Existing limits on light mediators}\label{ssec:yukawa_med_lim}

If there were no other constraints on $r_\phi$ or $\lambda_A$, Eq.~\eqref{mt_yukawa_highk} would allow  $\sigma^{\text{mt},(1)}_{\chi A}\gg10^{-25}\;\text{cm}^2$, albeit with a less useful scaling relation between different nuclei. However, because the light mediator couples to the Standard Model directly, other experiments already place constraints on such a particle. Figure~\ref{fig:combine_cross} shows the maximum achievable $\sigma_{\chi N}$ and $\sigma^{\text{mt}}_{\chi N}$ for a repulsive Yukawa potential, conservatively using the perturbativity limit $\lambda_\chi=4\pi$, $\mu_N\approx m_p$, and $k=0.005\;\text{fm}^{-1}$. When Eq.~\eqref{born_cond_approx} is $>10^{-4}$, Fig.~\ref{fig:combine_cross} uses the results from a numerical partial wave expansion with adaptive $l_\text{max}$, rather than the Born approximation. In practice, the Born approximation is adequate in the entire unconstrained region. 

Including all such constraints, we have $\sigma_{\chi N}\lesssim10^{-27}\;\text{cm}^2$ and $\sigma^{\text{mt}}_{\chi N}\lesssim10^{-32}\;\text{cm}^2$. Constraints that rely on lower relative velocities, such as the cosmological constraints discussed in Sec.~\ref{ssec:cosmology}, could achieve larger cross sections, but their constraints would need to be scaled correctly to compare them to direct-detection constraints. The momentum-transfer cross section is restricted to be $\sigma^{\text{mt}}_{\chi N}\lesssim10^{-25}\;\text{cm}^2$ even for velocities as low as $10^{-6}\,c$.

It is also possible to produce the light mediator in a collision \cite{Feng:2017uoz}. Particle production is an inelastic scattering process and beyond the scope of this paper, but it could be another avenue to transfer momentum between dark matter and a detector. 

The detailed constraints in Fig.~\ref{fig:combine_cross} could be different for different types of mediators. For example, for a vector mediator, the BBN constraints would be stronger \cite{light_mediator}. Other constraints might be weaker. However, $\sigma_{\chi N}\lesssim10^{-27}\;\text{cm}^2$ is already smaller than the geometric cross section of the nucleus, and circumventing individual constraints is unlikely to drastically change the overall conclusion that light mediators do not appear to be a promising approach to achieving large cross sections.

\section{Composite Dark Matter}\label{sec:composite_dm}
Another mechanism for achieving a larger characteristic interaction radius is dark matter that is not a point particle, but instead has a finite physical extent \cite{Kusenko:1997si,Khlopov:2005ew,Khlopov:2008ty,Krnjaic:2014xza,Detmold:2014qqa,Hochberg:2014kqa,Jacobs:2014yca,Hardy:2015boa,Garcia:2015loa,Hochberg:2015vrg,Farrar:2017ysn,DeLuca:2018mzn,Graham:2018efk,Hochberg:2018vdo,Grabowska:2018lnd,Coskuner:2018are}. Such dark matter could take the form of a composite particle. Because such dark matter would likely require an entire dark sector, any conclusions about the largest possible cross section with composite dark matter would be intrinsically model dependent. Because the largest physical scale in the problem is no longer related to a property of the target nucleus, the cross section need not scale with $A$ at all. 

The actual scaling with $A$ could only be determined by examining the particular model of composite dark matter. Additionally, achieving cross sections significantly larger than a nucleus with composite dark matter will always require $k r_{\text{dm}}\gtrsim 1$ for typical Milky Way virial velocities, so constraints on composite dark matter will need to be computed with a specific dark matter form factor in mind. See Sec.~\ref{ssec:cosmology} for a discussion of limits at the lower velocities relevant to cosmological limits. Analyses setting constraints on specific form factors at large cross sections should consider whether their specific choice of form factor can be achieved at the cross sections they are constraining in a physically realistic model. 

Therefore, limits on composite dark matter need to be calculated in specific models. Calculation of constraints on specific models of composite dark matter is left to future work. 


\section{Implications for Existing Constraints}\label{sec:implications}

\begin{figure}[tp]
    \includegraphics[width=\columnwidth]{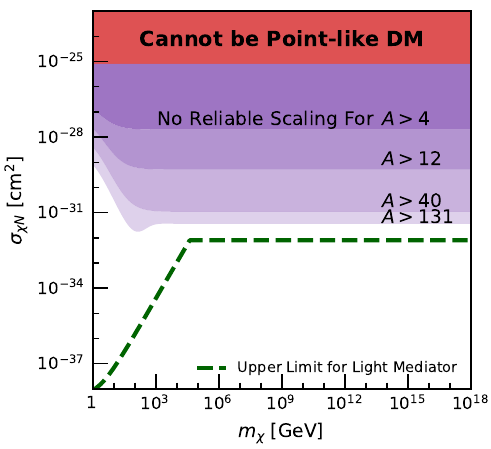} \\
    \caption{Summary of theoretically allowed regions for dark matter candidates. For a contact interaction, $A^4$ scaling breaks down for heavy nuclei for  $\sigma_{\chi N}\gtrsim10^{-32}\;\text{cm}^2$, and by $\sigma_{\chi N}\gtrsim4\times10^{-28}$ any scaling between different nuclei is model dependent. Here we define the failure of scaling as setting the lhs of Eq.~\eqref{final_cond_born_use} equal to $0.5$. This choice approximately agrees with where scaling obviously fails in Fig.~\ref{fig:sigma_sigma}. The breakdown is purely on theoretical grounds. Also shown is the maximum allowed momentum-transfer cross section for a $m_\phi=10^{-4}\;\text{GeV}$ light mediator using the constraints shown in Fig.~\ref{fig:combine_cross}, coincidentally at a comparable scale. For $m_\chi\lesssim10^4\;\text{GeV}$ we have applied a conservative self-interaction constraint $\sigma_{\chi\chi}/m_{\chi}<10\;\text{cm}^2/\text{g}$ \cite{Tulin:2017ara}. For $\sigma_{\chi N}\gtrsim10^{-25}\;\text{cm}^2$, no viable pointlike dark matter candidates exist.}
    \label{fig:scattering_scaling}
\end{figure}

Figure~\ref{fig:scattering_scaling} summarizes the approximate limits for the repulsive contact-interaction cross sections discussed in Sec.~\ref{sec:contact}. In the colored regions, the Born approximation begins to break down when the proton cross section is scaled to heavier nuclei, ultimately failing even for light nuclei. For pointlike dark matter with a contact interaction, cross sections much larger than the geometric cross section are completely forbidden. As discussed in Sec.~\ref{sec:light_mediator}, the limits for a light mediator are similarly below the geometric cross section. For $m_\chi\gtrsim 10^{16}\;\text{GeV}$, the entire (small) exclusion region for underground detectors is affected by the failure of scaling relations. Future improvements to constraints could change the region where the entire exclusion region would fail. Additionally, all detectors' computed ceilings are affected by the breakdown of scaling relations.

\subsection{Scaling constraints}
In the regime where scaling relations are unreliable, it becomes more difficult to compare constraints between experiments. When the scaling relations fail, scaling constraints from different nuclei to the dark matter-nucleon cross section using the $A^4$ is no longer meaningful.

For both contact interactions and light mediators, as the cross section begins to saturate, the momentum-transfer cross section scales less than linearly with $A$. Therefore, at fixed total detector mass, there is more detectable momentum transfer into the detector for \emph{lighter} target nuclei. The failure of the scaling relations also occurs at larger cross sections for smaller $A$. For example, a $^{12}$C-based detector would be able to use the Born approximation, and therefore the scaling relations, up to about 3000 times larger dark matter-nucleon cross section than a $^{131}$Xe-based detector. Therefore, robustly covering the large cross-section regime may be best accomplished by detectors using light nuclei, e.g.~\cite{Behnke:2016lsk,Amole:2017dex,Lippincott:2017yst,Collar:2018ydf}.

One option is to simply not scale constraints at large cross sections. While with resonances it could be possible for heavy nuclei to have smaller cross sections than a single nucleon, broadening by the dark matter velocity dispersion may limit the effect of narrow resonances on the overall detectable signature. Therefore, a relatively conservative approach could be to plot the actual momentum-transfer cross-section constraints obtained from different nuclei on the same scale. In fact, if composite dark matter as discussed in Sec.~\ref{sec:composite_dm} is indeed the most plausible strongly interacting dark matter candidate, disregarding scaling with $A$ may be the most correct way of plotting constraints. 
\subsection{Detection ceilings}
Now we briefly consider if the detection ceilings (i.e., the largest cross sections that can be probed by a given detector based on the detector's overburden) shown in Fig.~\ref{fig:prexist_limit_simpl} are preserved. In our simple model in Sec.~\ref{sec:contact}, cross sections simply saturate at 4 times the geometric cross section for heavier nuclei. Even if all nuclei in the detector overburden have an elastic scattering cross section equal to their geometric cross section, dark matter cannot be stopped by the overburden above some $m_\chi$ \cite{Goodman:1984dc}. 

Because all currently computed detector ceilings exist at cross sections where the breakdown of the $A^4$ scaling is severe, correctly calculated detector ceilings must be specialized to a specific model. For basic energy-independent cross-section scaling, the weakened ceilings likely lead to stronger direct-detection constraints for $m_\chi\lesssim 10^{16}\;\text{GeV}$. For such models, direct detection may even have exhausted the parameter space for cross sections up to the largest cross sections achievable with pointlike dark matter. For other dark matter form factors, the behavior around the ceiling could be more complicated.

Further work is required to make detailed adjustments to existing constraint contours to determine what dark matter parameter space has been constrained at large cross sections.
\begin{figure}[t]
    \includegraphics[width=\columnwidth]{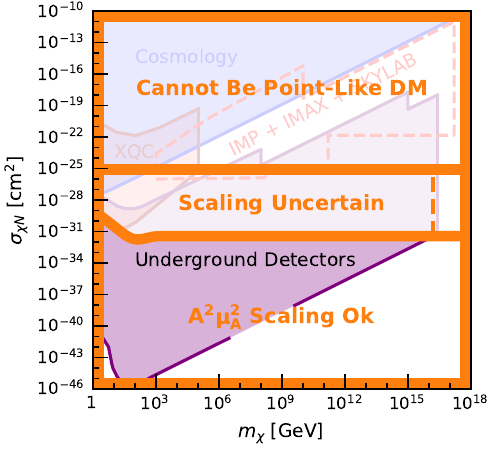} \\
    \caption{Claimed constraints from Fig.~\ref{fig:prexist_limit_simpl}, with the problematic regions identified in Fig.~\ref{fig:scattering_scaling} highlighted. All existing detector ceiling calculations are deeply in the model-dependent regime, or entirely excluded for pointlike dark matter. To the right of the dashed vertical line, the entire (small) direct-detection region must be reanalyzed.}

    \label{fig:limit_summary}
\end{figure}

\subsection{Dark matter-proton scattering constraints}\label{ssec:cosmology}

Constraints that rely only on dark matter scattering directly with protons are not directly affected by the breakdown of scaling relations with $A$. These are primarily constraints from cosmology and astrophysics, although at least one laboratory experiment uses proton targets \cite{Collar:2018ydf}. Astrophysics constraints (e.g. disk stability, stars, cosmic ray interactions, gas clouds, etc) are typically assumed to occur at galactic virial velocities, as for direct detection. Cosmology constraints, such as CMB and structure formation constraints, typically assume that collisions occur at smaller relative velocities. 

As shown in Fig.~\ref{fig:scattering_scaling}, the cross sections of interest for cosmological or astrophysical constraints are too large to be pointlike dark matter. Therefore, they should be reinterpreted as constraints on specific models of composite dark matter with a specified form factor, as discussed in Sec.~\ref{sec:composite_dm}. 

For cosmology constraints set at lower relative velocities, the suppression of the cross section by the form factor of dark matter is not as severe. One consequence is that it is possible to achieve somewhat larger cross sections for pointlike dark matter with a light mediator than those shown in Fig.~\ref{fig:combine_cross}, although even for velocities as low as $v\simeq 0.3\;\text{km/s}$, existing constraints would still require $\sigma^{\text{mt}}_{\chi N}\lesssim 10^{-25}\;\text{cm}^2$. However, invoking such a model would require additional caution, as direct-detection constraints would not be scaled correctly relative to the cosmology constraints, such that it would no longer be appropriate to plot cosmology and direct-detection constraints on the same axes, as done in Fig.~\ref{fig:prexist_limit_simpl}. 

Cosmological and astrophysical constraints set at masses $m_\chi<1\;\text{GeV}$, discussed in Sec.~\ref{ssec:low_mass_dm}, are at lower cross sections, and may still be meaningful constraints on pointlike dark matter. However, analyses at lower masses should either directly investigate how high their limits can be extrapolated, or make it much clearer that there are caveats in extrapolating their results to much larger masses.

\subsection{Low-mass dark matter}\label{ssec:low_mass_dm}
Because for $m_\chi\ll1\;\text{GeV}$, $\mu_A\simeq m_\chi$, low-mass dark matter constraints benefit only from a single factor of $A^2$ from coherence. Therefore, the loss of the $A^2$ scaling at large cross sections will be orders of magnitude less severe than the impact from the loss of $A^4$ scaling at larger masses. The momentum transfer is also smaller, so the loss of coherence due to an assumed form factor for the dark matter would be less severe. Contact interactions are still limited by the geometric size of the nucleus, but constraints on light mediators will become a function of $m_\chi$ \cite{light_mediator}. 

We leave a detailed assessment of the impact of our considerations at low mass to future work. However, we reiterate our caution that constraints set at low masses should carefully state the limitations on extrapolating their constraints to $m_\chi\gtrsim1\;\text{GeV}$.


\section{Conclusions}\label{sec:disc_conc}

How do dark matter particles interact with matter?  One of the most commonly considered cases to probe is the spin-independent interactions of $m_\chi > 1\;\text{GeV}$ pointlike dark matter with nuclei.  In the literature, a vast array of constraints --- based on astrophysical and cosmological tests, as well as direct-detection searches with a wide range of nuclei and overburdens --- are all compared to each other in simple plots of the dark matter-nucleon cross section and dark matter mass.  Comparing searches in this way requires the assumption of scaling relations, e.g., $\sigma_{\chi A} \propto A^4 \sigma_{\chi N}$ for $m_\chi \gg m_A$, that are widely assumed to be model independent.
 
We systematically examine the validity of the assumptions used to derive these relations, calculating where model independence ends.  Figure~\ref{fig:limit_summary} summarizes our results.  We find the following:
 
\begin{enumerate}
\item For small cross sections, $\sigma_{\chi N}\ll10^{-32}\;\text{cm}^2$, the usual scaling relations are valid, and multiple reasonable models can produce the same scaling relation.
\item For $10^{-32}\;\text{cm}^2\lesssim\sigma_{\chi N}\lesssim10^{-25}\;\text{cm}^2$, the assumed $A^4$ scaling for a contact interaction  progressively fails for all nuclear targets as cross sections for heavier nuclei begin to saturate at their geometric cross sections. Experimental constraints on the existence of light mediators prevent simple light mediator models from achieving cross sections in this range at all, such that constraints set in this range of cross sections should be specialized to a model. 
\item For $\sigma_{\chi N}>10^{-25}\;\text{cm}^2$, dark matter cannot be pointlike.  Contact interactions cannot achieve cross sections larger than the geometric cross section $\sigma_{\chi A}\simeq 4\pi r_A^2$, and simple light mediators are strongly ruled out. Dark matter with cross sections in this range must be composite.
\end{enumerate}
 
The failure of the scaling relations should influence the design of future dark matter searches. For interactions with cross sections that scale less than linearly with $A$, such as some models of composite dark matter, dark matter detectors with lighter nuclei are more efficient per unit detector mass. As a result, future direct-detection searches for strongly interacting dark matter may benefit from constructing detectors with light nuclei. 

Constraints on dark matter parameter space are most useful if they can be compared between different experiments. Where the $A^4$ scaling is not reliable, results need to be recast in terms of specific models. A comprehensive analysis should include clear statements about the mass ranges their results can reasonably be extrapolated to. Because constraints will not be the same for different models, plots including cross sections $\sigma_{\chi N}\gtrsim10^{-32}\;\text{cm}^2$ must specify a model, whether it involves a contact interaction, light mediator, composite dark matter, or something else. 


{\section*{Erratum}

This paper focused on calculating where model independence ends for the scaling relations relating the spin-independent elastic-scattering cross sections of dark matter ($m_\chi \gtrsim 1 \text{ GeV}$) between different nuclei.  We also made some broader statements about regions of parameter space that are ruled out on theoretical grounds for pointlike dark matter.  In the main text, these statements are accompanied by appropriate caveats; however, in some places --- mostly the the abstract, introduction, and conclusions --- the caveats were not always present in isolated passages.  In this erratum, we note some of those missing caveats more explicitly.  None of our main results or conclusions are affected.

An example of a type of statement that needs clarification is given in the Conclusions: \emph{``For $\sigma_{\chi N} > 10^{-25}\;\text{cm}^2$, dark matter cannot be pointlike. Contact interactions cannot achieve cross sections larger than the geometric cross section $\sigma_{\chi A} \simeq 4 \pi r_A^2$, and simple light mediators are strongly ruled out. Dark matter with cross sections in this range must be composite.''}  For repulsive interactions, this statement is correct. However, there are exceptions for attractive interactions, as shown in, e.g., Fig.~3, which shows resonances that can exceed the geometric limit. Though resonances are a generic feature of strongly attractive potentials (see, e.g., Refs.~\cite{regge65_scattering,Bai:2009cd, Xu:2020qjk}), the details of resonances are inherently model dependent, as discussed in Sec.~III~C.

When the incident momentum is small, as for very light nuclei or DM with $1\; \text{GeV} \lesssim m_{\chi} \lesssim m_A$, \emph{and} the potential is attractive, we acknowledge that the cross section can exceed the geometric limit by 3--5 orders of magnitude at typical halo velocities. We show this for helium in Fig.~3, which assumes $m_{\chi} \gtrsim m_A$. But when $m_{\chi} \gtrsim m_A$, no nuclei heavier than neon can have attractive cross sections even one order of magnitude larger than the geometric limit. The requirement of $m_{\chi} \gtrsim m_A$ is appropriate for nearly all of the mass range considered (e.g., the mass range in Fig.~7 spans eighteen orders of magnitude), but we should have been more clear than there are some subtleties in the narrow but important range where $m_{\chi} \lesssim m_A$.

For clarity, we also reiterate a point that was mentioned in the ``Overview of basic assumptions'' in Sec.~II~A but perhaps not made clear enough; that unless otherwise specified, we focus on Milky Way virial velocities, $v\sim 10^{-3}\;c$, relevant for direct detection. As we discuss in Sec.~VI~C, attractive s-wave cross sections could be even larger for velocities lower than $v\sim 10^{-3}\;c$, as is relevant for dwarf galaxies and especially for early universe cosmology. However, the limits on light mediator masses and couplings discussed in Sec.~IV~B do not depend on the interaction velocity and are so restrictive that, in the unconstrained regions, the upper limits on the cross section would essentially be the same for both attractive and repulsive interactions. For a model finely tuned to evade some of the light mediator constraints in Fig.~5, or an interaction potential generated by a different mechanism entirely, attractive interactions at lower relative velocities could potentially achieve much larger cross sections. Such cases would be model-dependent by construction.

We encourage further work on the many interesting possible model-dependent interactions affecting dark matter scattering with nuclei. We also reiterate that the scope of the paper is explicitly restricted to $m_\chi \gtrsim 1 \text{ GeV}$, so that we strongly discourage any attempt to extrapolate any of our results to lower masses without a careful supporting analysis. Such analyses will be an important complement to our work on where model independence ends. We thank Eric Braaten and Glennys Farrar for helpful discussions on these issues.}


\acknowledgements
We are grateful for useful discussions with Laura Baudis, Kimberly Boddy, Juan Collar, Adrienne Erickcek, Vera Gluscevic, Rafael Lang, Hitoshi Murayama, Ethan Nadler, and Juri Smirnov.  

MCD and CMH were supported by the Simons Foundation award 60052667, NASA award 15-WFIRST15-0008, and the US Department of Energy award DE-SC0019083. CVC and JFB are supported by NSF grant PHY-1714479 to JFB. AHGP is supported by NASA grant ATP 80NSSC18K1014 and NSF grant AST-1615838.


\appendix


\section{Lippmann-Schwinger Equation}\label{app:lippmann_schwinger}
It is useful to write $E=\frac{k^2}{2\mu}$, $U({\bf r})\equiv2\mu V({\bf r})$ and rearrange Eq.~\eqref{tise}
\begin{equation}\label{tise_green}
\left(\nabla^2_{\bf r}+k^2\right)\psi({\bf r}) =U({\bf r})\psi({\bf r}).
\end{equation}
Recognizing Eq.~\eqref{tise_green} as an inhomogeneous Helmholtz equation, we can write the general solution in integral form \cite{arfken,burke_joachain}:

\begin{equation}\label{tise_integral}
\psi({\bf r})=\psi_0({\bf r})+\int G_0\left({\bf r},{\bf r'}\right)U({\bf r'})\psi({\bf r'})d^3{\bf r'},
\end{equation}
where $G_0\left({\bf r},{\bf r'}\right)$ is the Green's function for an outgoing wave in the Helmholtz equation:
\begin{equation}\label{helmholtz_green_def}
\left(\nabla^2_{\bf r}+k^2\right)G_0\left({\bf r},{\bf r'}\right)=\delta({\bf r}-{\bf r'})
\end{equation}
and $\left(\nabla^2_{\bf r}+k^2\right)\psi_0({\bf r})=0$ is a homogeneous solution. The Green's function is given \cite{arfken} by
\begin{equation}\label{helmholtz_green_sol}
G_0\left({\bf r},{\bf r'}\right)=-\frac{1}{4\pi|{\bf r}-{\bf r'}|}e^{ik|{\bf r}-{\bf r'}|}.
\end{equation}
Plugging in an incident plane wave for the homogeneous solution, $\psi_0({\bf r})= (2\pi)^{-3/2}e^{i {\bf k}_i\cdot{\bf r}}$, where ${\bf k}_i\equiv k{\bf\hat{z}}$, we arrive at the Lippmann-Schwinger equation, 
\begin{equation}\label{lippmann_schwinger}
\psi({\bf r})=(2\pi)^{-3/2}e^{i {\bf k}_i\cdot{\bf r}}-\int U({\bf r'})\psi({\bf r'})\frac{e^{ik|{\bf r}-{\bf r'}|}}{4\pi|{\bf r}-{\bf r'}|}d^3{\bf r'}.
\end{equation}

Now, the goal here is to discover what measurable effect the potential has on the scattered wave. Physically, any measurement we make of the scattered wave must occur long after the particle has finished interacting with the potential. Therefore, we may safely assume $|{\bf r}|\gg|{\bf r'}|$, such that $|{\bf r}-{\bf r'}|\xrightarrow[]{r\rightarrow\infty}r-{\bf\hat{r}}\cdot{\bf r'}+\mathcal{O}(r^{-1})$, and defining ${\bf k}_f\equiv k{\bf\hat{r}}$, Eq.~\eqref{lippmann_schwinger} becomes 
\begin{align}
\psi({\bf r})\xrightarrow[]{r\rightarrow\infty}&\psi_0({\bf r})-\frac{e^{ikr}}{4\pi r} \int U({\bf r'})\psi({\bf r'})e^{-i{\bf k}_f\cdot{\bf r'}}d^3{\bf r'} \nonumber\\
\equiv&\psi_0({\bf r})+(2\pi)^{-3/2}\frac{e^{ikr}}{r}f\left({\bf k}_i,{\bf k}_f\right).\label{lippmann_schwinger_larger}
\end{align}
Physically, this equation represents an incoming plane wave and a radially outgoing spherical wave with scattering amplitude $f\left({\bf k}_i,{\bf k}_f\right)=f\left(k,\theta\right)$. 


\section{Born Approximation}\label{app:born_der}
Now we want to calculate an approximation to the scattering amplitude for a given potential. If we assume the potential is a perturbation to the incident wave function, we can attempt to solve Eq.~\eqref{lippmann_schwinger_larger} by iteration: 

\begin{align}
\psi({\bf r})\xrightarrow[]{r\rightarrow\infty}&\psi_0({\bf r})-\frac{e^{ikr}}{4\pi r} \int U({\bf r'})\psi({\bf r'})e^{-i{\bf k}_f\cdot{\bf r'}}d^3{\bf r'}\nonumber\\
=&\psi_0({\bf r})-\frac{e^{ikr}}{4\pi r} \int U({\bf r'})\left[\psi_0({\bf r'})-...\right]e^{-i{\bf k}_f\cdot{\bf r'}}d^3{\bf r'}\nonumber\\
=&\psi_0({\bf r})+(2\pi)^{-3/2}\frac{e^{ikr}}{r}(f^{(1)}\left({\bf k}_i,{\bf k}_f\right)+...),\label{lippmann_schwinger_iteration}
\end{align}
 where we have assumed the correction is small, such that higher-order corrections can be ignored. Then we can read off our approximation to $f\left({\bf k}_i,{\bf k}_f\right)$ from Eq.~\eqref{lippmann_schwinger}:
\begin{equation}\label{first_born_scattering}
f^{(1)}\left({\bf k}_i,{\bf k}_f\right)=-\frac{1}{4\pi}\int U({\bf r'})e^{i ({\bf k}_i-{\bf k}_f)\cdot{\bf r'}}d^3{\bf r'}.
\end{equation}
Here, $f^{(1)}\left({\bf k}_i,{\bf k}_f\right)$ is the first Born approximation to $f\left({\bf k}_i,{\bf k}_f\right)$. 

Inspecting Eq.~\eqref{first_born_scattering}, we recognize that the first Born approximation of $f\left({\bf k}_i,{\bf k}_f\right)$ is nothing more than the Fourier transform of the potential. Defining ${\bf q}\equiv{\bf k}_i-{\bf k}_f$ such that $q=|{\bf q}|=2k\sin\left(\frac{\theta}{2}\right)$, and assuming the potential to be spherically symmetric $U({\bf r'})=U({r'})$, we can perform the angular integration to obtain: 
\begin{equation}\label{first_born_fourier}
f^{(1)}\left({\bf k}_i,{\bf k}_f\right)=f\left(q\right)=-\frac{1}{q}\int_0^\infty U(r')\sin(qr')r' dr'.
\end{equation}
Equation~\eqref{first_born_fourier} is a useful starting point for analysis. However, before we begin using the result, we should clarify when the approximation breaks down. It can be shown robustly \cite{joachain75_collision} that a sufficient condition for the Born series to converge for all $k$ is that the magnitude of the potential would not be strong enough to support a bound state if it were purely attractive \cite{levinson1949uniqueness}, which is to say
\begin{equation}\label{born_convergence_guarantee}
\int_0^\infty r|U(r)|dr<1. 
\end{equation}
A useful heuristic condition for the validity of the first Born approximation at a given $k$ can be obtained by simply assuming the first-order correction term in Eq.~\eqref{lippmann_schwinger} must be small in the scattering region, such that $\psi({\bf r})\approx\psi_0({\bf r})$ near ${\bf r}=0$ \cite{sakurai,khare2012introduction}. Therefore, we require 
\begin{equation}
\frac{1}{4\pi}\left|\int U({\bf r'})e^{i {\bf k}_i\cdot{\bf r'}}\frac{e^{ik|{\bf r}-{\bf r'}|}}{|{\bf r}-{\bf r'}|}d^3{\bf r'}\right|\ll 1.
\end{equation}
Taking ${\bf r}=0$, replacing ${\bf k}_i\cdot{\bf r'}=kr'\cos\theta'$, and performing the angular integration, we have

\begin{equation}\label{final_cond_born}
\frac{1}{2k}\left|\int_0^\infty U(r')\left(e^{2ikr'}-1\right)dr'\right|\ll 1.
\end{equation}

Once we have the scattering amplitude, we can calculate the total cross section as in Eq.~\eqref{tot_scattering}.


\section{Partial wave analysis}\label{app:partial_wave}
The general scattering amplitude for a spherically symmetric potential in  Eq.~\eqref{lippmann_schwinger_larger} can be written as an arbitrary expansion of Legendre polynomials $P_l\left(\cos\theta\right)$ \cite{sakurai,joachain75_collision,burke_joachain}:
\begin{equation}\label{amplitude_partial}
f(k,\theta)=\frac{1}{k}\sum_{l=0}^\infty(2l+1)e^{i\delta_l}\sin\left(\delta_l\right) P_l\left(\cos\theta\right).
\end{equation}
Given the phase shifts $\delta_l(k)$, the total elastic scattering cross section can be readily evaluated:
\begin{equation}\label{partial_wave_cross_section}
\sigma^{\text{tot}}=\frac{4\pi}{k^2}\sum_{l=0}^\infty{(2l+1)\sin^2\left(\delta_l\right)}.
\end{equation}

A spherically symmetric wave function can be written as a linear combination of Bessel functions of the first and second kind, $j_l(k r)$ and $n_l(k r)$. The scattered part of the wave function for $r>R$, where $R$ is some arbitrarily large cutoff radius for the potential, can then also be expanded in terms of Legendre polynomials:
\begin{equation}\label{scattered_wave}
\psi_{\text{scattered}}(r,\theta)=\sum_{l=0}^\infty{A_l(r)P_l(\cos\theta)},
\end{equation}
where 
\begin{equation}
A_l(r)=e^{i\delta_l}\left[\cos\left(\delta_l\right)j_l(k r)-\sin\left(\delta_l\right) n_l(k r)\right]
\end{equation}
is the radial wave function for the $l$th partial wave. We can obtain $\delta_l(k)$ by enforcing continuity of the logarithmic derivative of the wave function,
\begin{equation}\label{wf_derivative}
\beta_l=\frac{r}{A_l}\frac{d A_l}{d r},
\end{equation}
at $r=R$. We can obtain the wave function for $r<R$ by directly integrating the one-dimensional Schrödinger equation,
\begin{equation}\label{1d_schro}
\frac{d^2u_l}{d r^2}+\left(k^2-2\mu V(r)-\frac{l(l+1)}{r^2}\right)u_l(r)=0,
\end{equation}
where we have defined $u_l(r)\equiv r A_l(r)$. Note that because we are matching the phase shift with the logarithmic derivative of the wave function, the overall normalization of $A_l(r)$ is irrelevant for our purposes and can be chosen arbitrarily. We can then obtain $A_l(R)$ for any arbitrary potential $V(r)$ by analytically or numerically evaluating Eq.~\eqref{1d_schro} from $r=0$ to $r=R$.

Once we have $A_l(R)$, we can obtain $\delta_l$ using
\begin{equation}\label{delta_l}
\tan(\delta_l)=\frac{k R A_l j'_l(k R)-(d A_l/d \ln{r})|_{r=R}j_l(k R)}{k R A_l n'_l(k R)-(d A_l/d \ln{r})|_{r=R}n_l(k R)}.
\end{equation}
We have avoided canceling $A_l(r)$ in order to preserve signs, to ensure we obtain the correct quadrant for $\delta_l(k)$.

The boundary condition at $r=0$ should properly be $u_l(0)=0$, but for numerical solutions taking the boundary to be at $r=0$ is inconvenient because of the $1/r^2$ centrifugal term. Instead we can take advantage of the arbitrary normalization of $A_l(r)$ and fix the boundary conditions $u_l(r_{\text{min}})=1$, $d u_l/dr(r_{\text{min}})=(l+1)/r_{\text{min}}$, where $r_{\text{min}}$ is some small minimum radius. 

As $k\rightarrow0$, it can be shown \cite{joachain75_collision} that generically $\delta_l(k)\propto k^{2l+1}$, except in special cases where it is possible to achieve $\delta_l(k)\propto k^{2l-1}$ for a specific value of $l$. Inspecting Eq.~\eqref{partial_wave_cross_section}, we can see the contributions from $l=0$ and $l=1$ are the only values of $l$ which can be nonvanishing as $k\rightarrow0$. The $l=0$ cross section is called the s-wave cross section.

\bibliographystyle{JHEP}
\bibliography{main}
\end{document}